\documentclass[preprint,10pt]{elsarticle}

 \usepackage{graphicx}
\usepackage{amssymb}
\usepackage[superscript,biblabel]{cite}

\usepackage{amsmath}

\usepackage{helvet}

\journal{OJMS}

\begin{document}
\baselineskip15pt
\noindent\textsc{Original research}\\
Simultaneous growth of two cancer cell lines\\
Hamon et al
\vskip 1cm\noindent
{\bf{\large Simultaneous growth of two cancer cell lines evidences
  variability in growth rates}}
\vskip 1cm\noindent
Agn\`es Hamon\textsuperscript{1,2,3}\\
Marie Tosolini\textsuperscript{3,4,5,6}\\
Bernard Ycart\textsuperscript{1,2,3}\\
Fr\'ed\'eric Pont\textsuperscript{3,4,5,6}\\
Jean-Jacques Fourni\'e\textsuperscript{3,4,5,6}
\vskip 1cm\noindent
\textsuperscript{1}{Universit\'e Grenoble-Alpes, France}\\
\textsuperscript{2}{Laboratoire Jean Kuntzmann, CNRS UMR 5224, Grenoble, France}\\
\textsuperscript{3}{Laboratoire d'Excellence `TOUCAN', France}\\
\textsuperscript{4}{INSERM UMR1037, Cancer Research Center of
  Toulouse, Toulouse, France}\\
\textsuperscript{5}{Universit\'e Toulouse III Paul-Sabatier, Toulouse, France}\\ 
\textsuperscript{6}{ERL 5294 CNRS, Toulouse, France}
\vskip 1cm\noindent
Correspondence: Bernard Ycart\\
51 rue des math\'ematiques, 38041
  Grenoble cedex, France\\
Tel: +33476514995\\
Fax: +33476631263\\
E-mail: bernard.ycart@imag.fr
\newpage
\noindent{\bf Abstract:}
Cancer cells co-cultured in vitro reveal unexpected differential growth
rates that classical exponential growth models cannot account for. 
Two non-interacting cell lines were grown 
in the same culture, and counts of each species were recorded at periodic
times. The relative growth of population ratios
was found to depend on the initial
proportion, in contradiction with the traditional exponential growth
model. The proposed explanation is the variability of growth rates for
clones inside the same cell line. This leads to a log-quadratic growth
model that provides both a theoretical explanation to the
phenomenon that was observed, and a better fit to our growth data.

%\end{frontmatter}
% \linenumbers
%double space
\vskip 5mm\noindent
\section*{Introduction}
\label{intro}
\begin{figure}
\hspace{-0.2cm}
\includegraphics[width=12cm]{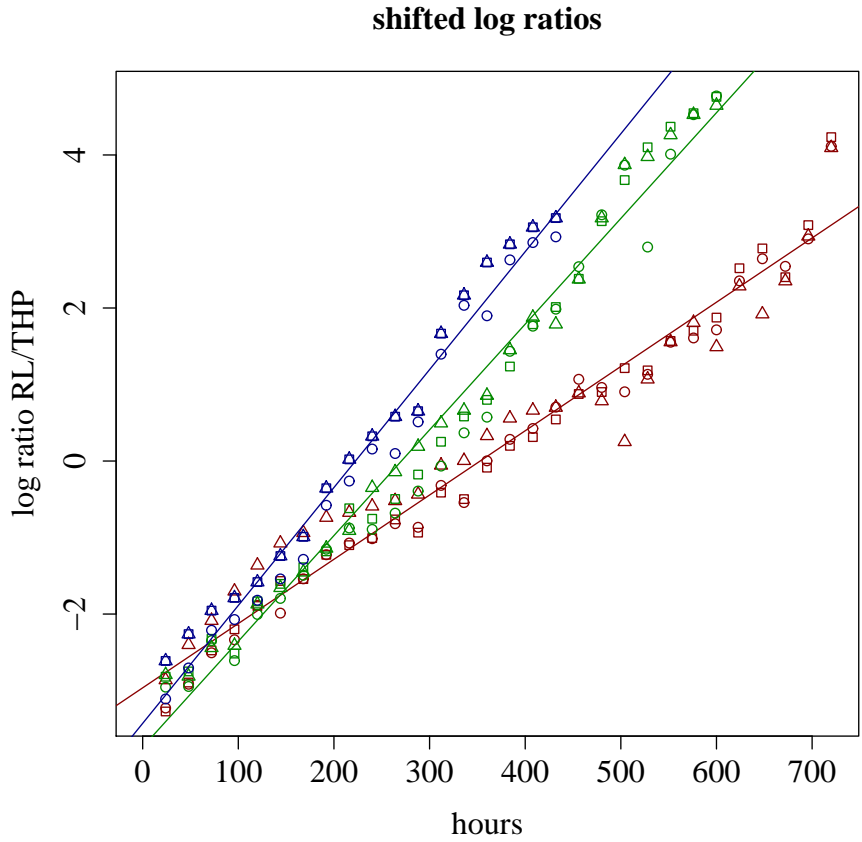}
\caption{Logarithms of ratios of RL/THP-1. The three replicates for
  each initial condition are 
  marked by circles, triangles and squares. All curves have been
  shifted to the first day where the proportion RL/THP-1 passes $5\%$.
Red marks correspond to an
  initial proportion of $5\%$, green marks to $1\%$, blue marks
  to $0.5\%$. The three regression lines are represented with
  corresponding colors. The slopes (time unit: hour) 
are $0.0084$ (red line, initial
  proportion $5\%$), $0.0138$ (green line, initial
  proportion $1\%$), and $0.0154$ (blue line, initial
  proportion $0.5\%$).}
\label{fig1}
\end{figure}

Since emergence of resistant cells underlies time to relapse for
cancer patients undergoing chemotherapy, the growth rate of these
tumor cells is a  crucial issue. Cancer therapies usually yield
undetectable levels of residual and resistant cancer stem cells (CSC)
in patients. Upon repeated mitosis however, CSC can seed a cell
progeny that progressively reconstitutes tumors, but the proportion
and mitotic rate of such CSC are highly variable in treated
patients. 
The classical exponential growth
model predicts that the relative growth of the fast-growing clones
should increase exponentially with time, regardless of their initial
rates in patients. On the other hand however, this model is challenged
by heterogeneity of the clonal progeny from a cancer cell and the
resulting Darwinian selection in this progeny for access to
nutriments\cite{Chevin11,Mannaetal12}. 
To investigate this, we grew either separately or together,
two different non-interacting human cancer cell lines, in cell
cultures with unlimited medium supply, and modeled the cell growth
rates observed in the co-cultures. 
The exponential growth model is so elementary and has been known
for such a long time\cite{Powell56,KimmelAxelrod02}, 
that it seems almost too simple to actually fit real
cell growth data\cite{KoutsoumanisLianou13}. 
Yet, for a given cell line grown in unlimited supporting medium, 
an excellent linear fit is usually observed for 
the logarithm of population size against 
time\cite{Deenicketal03,Zilmanetal10}. Our
experiments were conducted with two well known lab strains: 
RL (non-Hodgkin's lymphoma B cell line:
 ATCC CRL-2261\cite{Beckwithetal90}) and
THP-1 (cell line derived from an acute monocytic leukemia patient:
TCC TIB-202\cite{Tsuchiyaetal80}). 
As a control experiment, the two cell lines were grown separately. 
An excellent
log-linear fit was observed. Then, both strains were grown in the same
solution. No interaction between the two species could occur, other
than possible competition for nutrient from culture medium. 
This was avoided again
by maintaining a sufficient supply of medium by volume, and continuous
renewal.  
The initial proportions of
the faster growing strain RL were fixed at $0.5\%$, $1\%$, $5\%$, and
three replicates were made for each initial proportion. With the
classical exponential growth model, the relative proportion of RL \emph{vs.}
THP-1 would be predicted to increase exponentially in time,  at a rate
which is independent from the initial proportion. 
Somewhat unexpectedly, this turned out to be false. 
Figure \ref{fig1} presents a plot of the ratio of observed 
RL vs THP-1 counts 
on a logarithmic scale. The time scales have been shifted so that the origin
corresponds to the time at which each proportion reaches $5\%$.  
The slope of the regression line decreases as the
initial proportion of the faster strain increases: the slope 
with an initial proportion $5\%$ (red) is
smaller than that with initial proportions $1\%$ (green) and
$0.5\%$ (blue).

In vitro experiments with simulateous growth of two or more
 microorganisms have long been carried through: see Dykhuizen
 \cite{Dykhuizen90}
for a review. The variability of growth rates in human leukemia cell
clones
has been studied by Tomelleri et al\cite{Tomellerietal08}.
However, to the best of our knowledge, this is the first
instance describing an experiment with two different cancer
lines, and showing the phenomenon of 
dependence of the ratio growth rate on the initial condition,
illustrated on Figure \ref{fig1}.  

The objective of this paper is to propose a stochastic growth
model explaining the phenomenon, 
and show that the fit of the data by that model
is better than with the exponential growth model.
\vskip 2mm
It has long been known that exponential proliferation is a valid
approximation, only on a certain fraction of the observation
time\cite{Novick55}. Many different models have been proposed as
growth 
curves\cite{Zwieteringetal90,PelegCorradini11}. 
At the beginning of a cell growth experiment, 
a lag phase\cite{Brook81,Delignette98,BatyDelignette04} is usually observed;
this is the case in our data.
The lag phase 
could partly account for the phenomenon investigated here. Indeed,
when starting from a $0.5\%$ proportion, the lag phase has elapsed
when reaching $1\%$, but if one already starts with a $1\%$
proportion, the lag phase only begins. 
Yet the lag phase does not explain differential growth
rates after all cultures have reached a $5\%$ proportion, since at
that time, the lag
phase has elapsed in the first two cases.  
Another simple explanation is proposed here: 
intrinsic variability of growth 
rates\cite{KoutsoumanisLianou13,Tomellerietal08}.
Here, the notion of growth rate is understood in the
  sense of branching processes\cite{KimmelAxelrod02}, as a 
``large scale approximation'' that applies to the
whole clone stemming from a given cell, and not just to that
cell. Therefore, the variation of
growth rates can only be genetic: to each cell present at the
beginning of the experiment is associated one value, which will be
the growth rate of the whole clone stemming from that cell. 
If growth rates among RL vary, the
proportion of fast dividing mitotic cells among RL cell clones
will  gradually increase. When
reaching the proportion of $5\%$, there will be more fast breeders
among RL if the initial proportion was $0.5\%$, than if it was
$1\%$. This intuitively explains why the estimated growth
  rate over a given time interval, is larger when the proportion of RL
  reaches $5\%$, after starting from $0.5\%$. 

\begin{figure}
\hspace{-0.2cm}
\includegraphics[width=12cm]{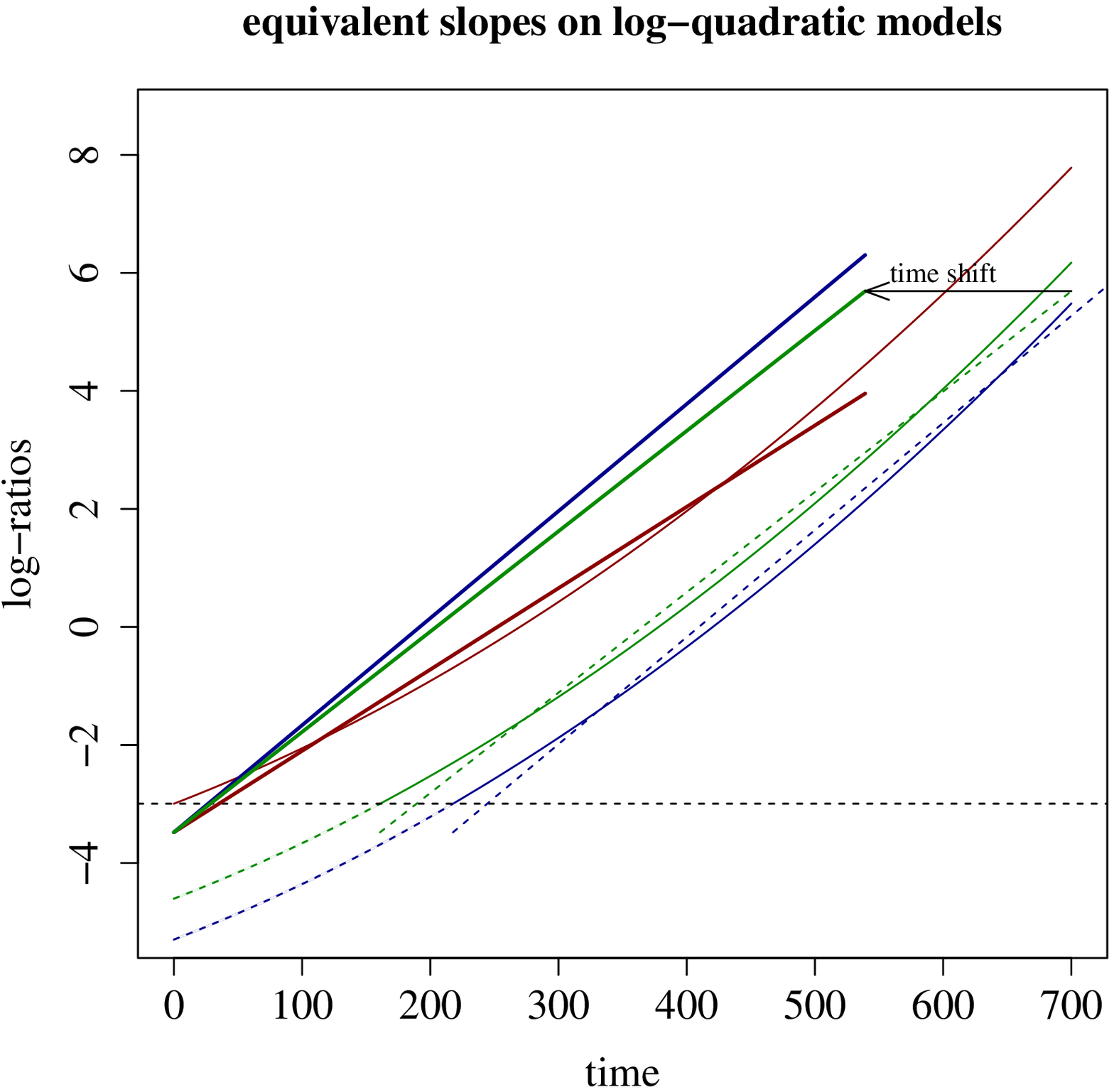}
\caption{Theoretical explanation for observed
differential growth rates. The
  figure represents a quadratic growth in time, at 3 different
  intercepts: $\log(0.05)$ (red), $\log(0.01)$ (green), $\log(0.005)$
  (blue). The three solid curves are parabolas. The green and blue dashed
  lines are linear fits over an interval starting at the point where
  the corresponding parabola reaches $\log(0.05)$. The green and blue solid
  lines are time shifts, illustrating the differential slopes.}
\label{fig2}
\end{figure}

Mathematically, it will be shown 
that assuming variable growth rates among the cells of
a same species, naturally leads to a log-quadratic model on the
population growth, instead of the traditional log-linear (or exponential)
model. It will be shown that the log-quadratic model
induces a better fit on our data. 
Using that model, the observed phenomenon can be explained and
quantified. Indeed, if the actual growth is log-quadratic instead
  of log-linear, the fit of a log-linear model yields estimated slopes
  that vary 
  with the initial condition: Figure \ref{fig2} illustrates that
  theoretical explanation. The derivation of the
log-quadratic model from the hypothesis of variable growth rates uses
the cumulant generating function of the random growth rates. A similar
explanation had been proposed by Hansen\cite{Hansen92}.

\vskip 5mm\noindent
\section*{Material and methods}

\subsection*{Experimental methods}
The cell line THP-1 (ATCC TIB-202) derives from human acute monocytic leukemia.
It has a monocyte morphology and expresses the cell
surface marker CD13. 
The cell line RL (ATCC CRL-2261) was derived from human non-Hodgkin's
lymphoma and expresses the cell surface marker CD20. 
These two cell lines were cultured as indicated by the supplier (ATCC
www.lgcstandards-atcc.org) at 37\textsuperscript{o}C and 5\% 
CO2 in liquid medium
 RPMI-1640 (LONZA, Levallois, France) supplemented with 10\%
 heat-inactivated fetal calf serum
 (FCS), 2mM L-glutamine, 100 U/ml
penicillin and 100µg/ml streptomycin
(Invitrogen, Cergy Pontoise, France). This medium contains
inorganic salt, amino acid, vitamins
and D-glucose (2g/L). 
THP-1 and RL, alone and in competition, were cultured in T75 flasks 
with 50mL of medium without agitation. 
At the beginning of the culture, for all the conditions, cells were
seeded at $0.3\times 10^6$ cells/mL. 
Daily, the differents cell cultures were counted and if necessary,
diluted with complete medium if the cell concentration was higher than
$0.7 \times 10^6$ cells/mL to adjust at $0.3 \times 10^6$ cells/mL. 
These concentrations
lead to no competition for nutrient. 
The dilutions of the culture were adjusted to the cell growth in each
flask: RL cell culture was more diluted than THP-1 cell culture for
example.
Cells from competition cultures were also analysed by flow cytometry
to determine the percentage of each cell line in the culture: Cells
were centrifuged, washed with PBS and incubated 10 min with antibodies
against CD20 coupled with the fluorochrome APC-Cy7 to identify RL
cells, and against CD13 coupled with the fluorochrome PE (BD
Biosciences, Pont de Claix, France) to identify THP-1 cells. 
Cells from single cell line culture were used as controls.
The fluorescence of 50000 cells was then analysed using a BD LSR II
cytometer. 

Two sets of experiments were carried through 35 days, with daily
measurements. 
In the first set, the two cell lines
RL and THP-1 were grown in separate culture flasks, in duplicate. 
In the second set, the two
cell lines were grown in the same culture flask, in triplicate, 
with unlimited amount of nutrient in each case. Three initial
proportions of RL (the faster growing strain) were considered: 
$0.5\%$, $1\%$, and $5\%$. 
For each set of experiment, each day of culture, and each replicate,
numbers of cells of each type were recorded. The dataset is available
upon request. On these data, different least square fits of the
log-quadratic model (\ref{quadratic}) were performed, 
for each of the two separate growths (first set of experiments), and
for simultaneous growth (second set).
\subsection*{Mathematical model}
In this section, a mathematical derivation of a log-qua\-dra\-tic growth
model, based on variable growth rates, is proposed.
 
Consider first the classical model of exponential growth for a single clone,
stemming from the general theory of 
branching processes\cite{Harris63,AthreyaNey72,KimmelAxelrod02}. 
From a single cell at
time $0$, the clone grows to size $N(t)$ at time $t$. Under fairly
general hypotheses on the division time distribution, there exists a
positive constant $b$, the growth rate (also called Malthusian
  parameter), such that almost surely:
\begin{equation}
\label{branching}
\lim_{t\to+\infty} \mathrm{e}^{-b t} N(t)= C\;,
\end{equation}
where $C$ is a random variable with finite expectation and
variance. This is one of the
basic results of the theory of branching
processes\cite{Harris63,AthreyaNey72}. 
Thus it is reasonable to assume
$N(t)=C\mathrm{e}^{b t}$ as a model of growth curve for a single
clone. Assume now that the population grows
from a large number $n$ of identical initial cells. For
$i=1,\ldots,n$, let $N_i(t)$ be
the size at time $t$ of the clone stemming from cell $i$:
\begin{equation}
\label{defNi}
N_i(t) = \mathrm{e}^{b t} C_i\;,
\end{equation}
where the $C_i$ are independent identically distributed random variables.
 The total
population at time $t$ is:
$$
N(t) = \sum_{i=1}^n N_i(t)\;.
$$
By the law
of large numbers, almost surely: 
$$
\lim_{n\to\infty} \frac{N(t)}{n} = \mathrm{e}^{\nu t} \mathbb{E}(C)\;,   
$$
where $\mathbb{E}(C)$ denotes the mathematical expectation of the
random variable $C$.
This justifies the classical log-linear model:
\begin{equation}
\label{linear}
\log(N(t)) = a+b t\;,
\end{equation}
where $a=\log(N(0))$. 
General references on log-linear models are
Mair\cite{Mair06} and von Eye and Mun\cite{vonEyeMun13}.

Consider now a second population growing according to the same model,
and denote by $M(t)$ its size at time $t$.
$$
\log(M(t)) = a'+b' t\;.
$$
Assume $b>b'$ (the first population grows faster). Then the ratio
$R(t)=N(t)/M(t)$ also follows a log-linear model.
$$
\log\left(R(t)\right) = (a-a')+(b-b') t\;.
$$
Whatever the interval of time it is observed on, 
the growth rate $b-b'$ does not depend on the interval nor on the
initial proportion.
 This is contradicted by our
observation (Figure \ref{fig1}).

Assume now that clones stemming from different initial cells may have
different growth rates. The new model is:
\begin{equation}
\label{defNi2}
N_i(t) = \mathrm{e}^{B_i t} C_i\;,
\end{equation}
where $(B_i,C_i)$ are independent and 
identically distributed copies of a random couple $(B,C)$. 
The joint distribution of
$(B,C)$ is of course unknown, and we shall make the technical
assumptions that $B$ and $C$ are independent, and that $B$ has
faster than exponential decaying tails. By the same argument of law of
large numbers, the global population $N(t)$ should satisfy:
$$
\log(N(t))= a+\log\left(\mathbb{E}\left(\mathrm{e}^{B t}\right)\right)\;.
$$
Note that the function 
$\log\left(\mathbb{E}\left(\mathrm{e}^{B t}\right)\right)$ 
exists for all $t\geqslant 0$
if the distribution of $B$ has faster than exponential decaying
tails. This function is the cumulant generating function of 
$B$\cite{KendallStuart94}, well known in 
large deviation theory\cite{DemboZeitouni98}. 
Let $\mu$ be the expectation of $B$, $\sigma$ its standard deviation
and $\gamma_1$ its skewness. Then the first three terms of the Taylor
expansion of 
$\log\left(\mathbb{E}\left(\mathrm{e}^{B t}\right)\right)$ are: 
\begin{equation}
\label{cgf}
\log\left(\mathbb{E}[\mathrm{e}^{B t}]\right) = \mu\,t+ \frac{\sigma^2}{2}\, t^2+
\frac{\sigma^3 \gamma_1}{6}\,t^3+o(t^3)\;,
\end{equation}
In the particular case where $B$
follows the Gaussian distribution, the first two terms give the exact
expression:
$$
\log(\mathbb{E}[\mathrm{e}^{B t}]) = \mu\,t+ \frac{\sigma^2}{2}\, t^2\;.
$$
In that case, 
the growth of $N(t)$ is quadratic in logarithmic scale:
\begin{equation}
\label{quadratic}
\log(N(t)) = a+b t+ct^2\;,
\end{equation}
with $a=\mathbb{E}[N_0]$, $b=\mu$, and $c=\sigma^2/2$.
Equation (\ref{quadratic})  will be referred to as \emph{log-quadratic
  model}: see Chapter 9 of von Eye and Mun\cite{vonEyeMun13}, and 
Stone et al\cite{Stoneetal09} for an application in a similar context. 
Assuming that the distribution of $B$ is
Gaussian may seem unrealistic, but whatever the distribution of $B$,
if its expectation is $\mu=b$ and variance $\sigma^2=2c$, equation
(\ref{quadratic}) remains true as a second order approximation,
because of (\ref{cgf}). This justifies the use of
(\ref{quadratic}) as a model, in case of variable growth rates.

If two populations grow according to a log-quadratic
model, then the ratio of the two population sizes does too. Denote
again that ratio by $R(t)$, assuming that the choice has been made to
put the faster growing population on the numerator, so that $R(t)$
increases.
\begin{equation}
\label{Rlq}
\log(R(t))=a+b t +c t^2\;,
\end{equation}
with $a=\log(R(0))$ and $b,c>0$.
In practice, growth rates are
estimated by a log-linear regression over a given interval, 
say $[T_1,T_2]$. This
amounts to approximating 
(\ref{Rlq}) by: 
$$
\log(R(t)) = \hat{\alpha}+\hat{\beta} t\;,
$$  
where $\hat{\alpha}$ and $\hat{\beta}$ are optimal in the sense of mean squares:
\begin{equation}
\label{minsquare}
(\hat{\alpha},\hat{\beta}) = \arg\min \int_{T_1}^{T_2} (a+bt+ct^2-\alpha-\beta
t)^2\,\mathrm{d} t\;.
\end{equation}
The solution of (\ref{minsquare}) is easily obtained:
$$
\hat{\alpha} = a-\frac{c}{6} (T_1^2+T_2^2+4T_1T_2)
\quad\mbox{and}\quad
\hat{\beta} = b+c (T_1+T_2)\;.
$$
For a fixed span $T_2-T_1$,
the ``equivalent growth rate'' $\hat{\beta}$ 
increases as  $T_1$ increases  (see
  Figure \ref{fig2} for an illustration).
This explains the
phenomenon evidenced by Figure \ref{fig1}. More precisely, let $T_1$
be the time at which $R$ reaches the value $R_1>R(0)$:
$$
T_1 = \frac{1}{2}\left( -b +\sqrt{b^2 
+4c\log\left(\frac{R_1}{R(0)}\right)}\right)\;.
$$
The equivalent growth rate on a time interval of duration $t$ 
after $T_1$ will be  $\hat{\beta}=b+c(2T_1+t)$. It will be larger than the
growth rate on an interval of same width starting at $0$, which is 
$b+c t$.

Thus the log-quadratic model (\ref{quadratic}) 
provides a theoretical explanation for the phenomenon of
differential growth rates, that has been observed. As will be shown in the
next section, it also provides a better fit to our data.

\vskip 5mm\noindent
\section*{Results}
\subsection*{Separate growth of RL}
\begin{figure}
\hspace{-0.2cm}
\includegraphics[width=12cm]{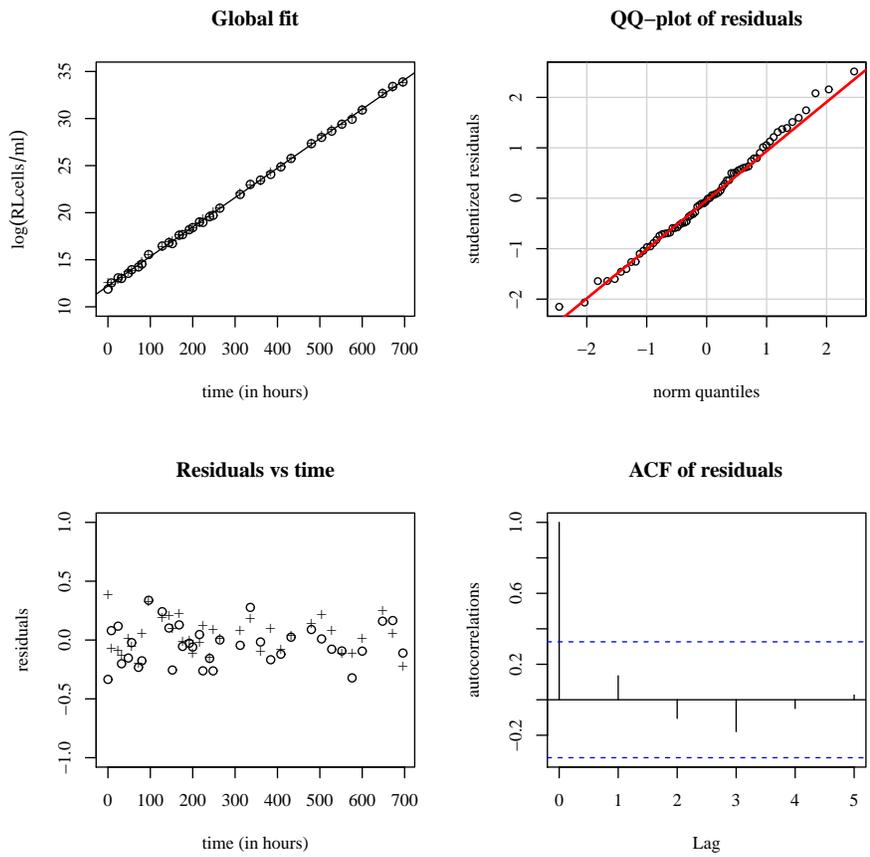}
\caption{Fit and validation of the log-linear model: RL cells.}
\label{res_RL}
\end{figure}

Let $Y_{ik}$ denote the logarithm of cell count at time $t_k$ and
replicate $i$ ($i=1,2$). 
We consider the following model :
\begin{equation}
\label{modelseparate}
Y_{ik}=a+bt_k+ct_k^2+\varepsilon_{ik}\;,
\end{equation}
where $\varepsilon_{ik}$ are centered Gaussian random variables with
common standard-deviation. For RL cells, it turned out that 
the coefficient $c$, that we shall call ``curvature'', 
was not significantly different from zero ($P=0.698$). 
Therefore a linear model without quadratic term was
fitted. Table \ref{estim_RL} reports the estimated coefficients. Figure 
\ref{res_RL} presents the residual analysis. The two coefficients $a$
and $b$ are significantly different from zero. 
The $95\%$ confidence interval
of the mean RL growth rate $b$ is $[0.0311;0.0315]$. This corresponds
to a doubling time between $22$ and $22.3$ hours.  
The proportion of the variation of $Y_{ik}$ explained 
by the fitted model is excellent ($R^2=0.99$).
The QQ-plot of residuals (figure
\ref{res_RL}) is close to linear, the plot of residuals \emph{vs.}
time does not show any mis-specification of the non-random part nor
heteroscedasticity problem. The Durbin-Watson test ($P=0.12$)
and the runs test ($P=0.69$) indicate no violation of the
hypothesis of error independence.  

\begin{table}
\begin{center}
\begin{tabular}{rccc}\hline
Coefficient&Estimate & Std. Error & $P$\\ \hline   
$a$ & $12.21$ & $3\times 10^{-2}$ & $ <2\times 10^{-16}$\\
$b$ & $0.0313$ & $9\times 10^{-5}$ & $<2\times 10^{-16}$\\\hline
\end{tabular}
\caption{Estimations for the logarithm of RL cell counts.}
\label{estim_RL}
\end{center}
\end{table}

\subsection*{Separate growth of THP-1}
\begin{figure}
\hspace{-0.2cm}
\includegraphics[width=12cm]{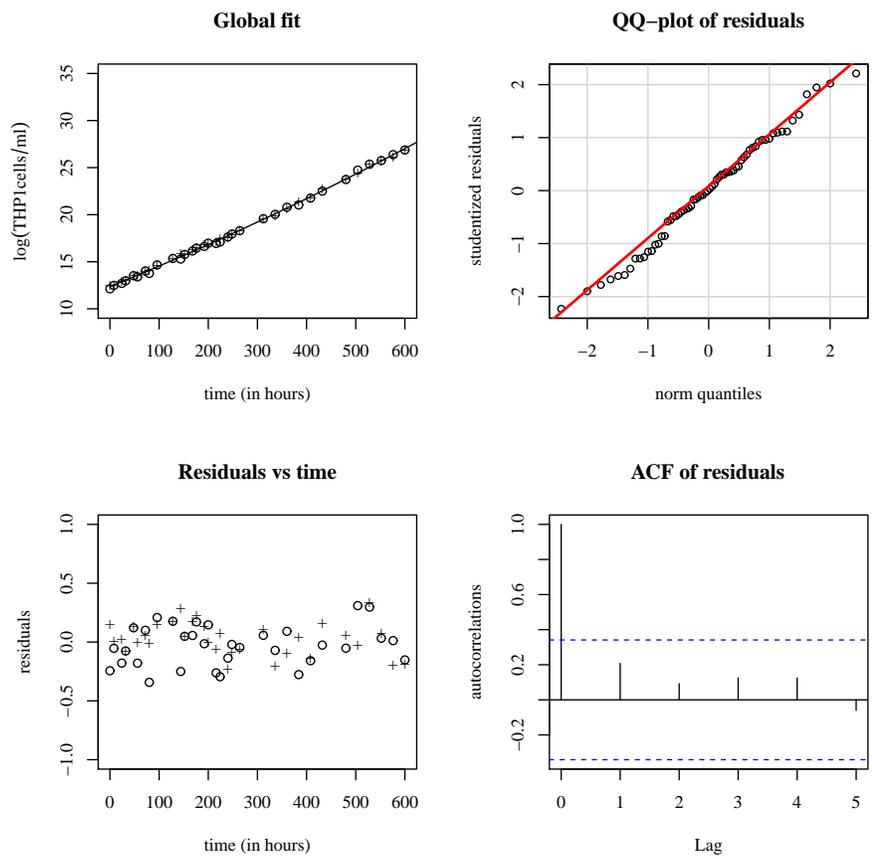}
\caption{Fit and validation of the log-quadratic model: THP-1 cells.}
\label{res_THP1}
\end{figure}

The model remains the same, see equation (\ref{modelseparate}).
At first, it was fitted to the full data set. Three observations
at the end of the experiment period, were detected as outliers, and
therefore excluded from the final analysis. 
Table \ref{estim_THP1} reports the estimated coefficients. Figure 
\ref{res_THP1} presents the residual analysis. Contrarily to the RL case, 
the curvature $c$ is significantly positive ($P=8.6\times 10^{-11}$).
The proportion
of the variation of $Y_{ik}$ explained by the
fitted model is excellent ($R^2=0.99$).  
The $95\%$ confidence interval of the mean THP-1 growth
rate $b$ is $[0.0204;0.0227]$. This  corresponds to a doubling time
between $30.5$ and $34$ hours, i.e. slightly below 
the values given in Tsuchiya et al\cite{Tsuchiyaetal80}
(35 to 50 hours), and above those of Tsuchiya et 
al\cite{Tsuchiyaetal86} (24 to 30 hours).

The QQ-plot of residuals (figure
\ref{res_THP1}) is close to linear, the plot of residuals \emph{vs.}
time does not show any mis-specification of the non-random part nor
heteroscedasticity problem. 
The Durbin-Watson test
($P=0.16$) and the runs test ($P=0.75$)
indicate no violation of the hypothesis of error independence.

As expected, 
the growth rate of THP-1 is significantly
smaller than that of RL ($P<0.0001$).

\begin{table}
\begin{center}
\begin{tabular}{rccc}\hline
Coefficient&Estimate & Std. Error & $P$\\ \hline   
$a$ & $12.37$ & $5\times 10^{-2}$ & $ <2\times 10^{-16}$\\
$b$ & $0.0212$ & $4\times 10^{-4}$ & $<2\times 10^{-16}$\\
$c$ & $5\times 10^{-6}$ & $7\times 10^{-7}$& $8.6\times 10^{-11}$\\\hline
\end{tabular}
\caption{Estimations for the logarithm of THP-1 cell counts.}
\label{estim_THP1}
\end{center}
\end{table}

\subsection*{Simultaneous growth of RL and THP-1}
Let $Y_{ijk}$ denote the logarithm of the ratio of RL by THP-1 cell
counts at time $t_k$, where $i$ denotes the replicate ($i=1,2,3$) and 
$j$ the initial nominal value of the ratio, that will be called
\emph{dilution}. Indices 
$j=1,2,3$ correspond to dilutions $0.5\%,1\%,5\%$.
Four different models were considered: 
\begin{align}
\tag{M0}
Y_{ijk} &= a +a_j+b t_k+\epsilon_{ijk}\\
\tag{M1}
Y_{ijk} &= a +a_j+b t_k+b_j t_k+\epsilon_{ijk}\\
\tag{M2b}
Y_{ijk} &= a +a_j+b t_k+b_j t_k+c t_k^2+\epsilon_{ijk}\\
\tag{M2c}
Y_{ijk} &= a +a_j+b t_k+c t_k^2+c_j t_k^2+\epsilon_{ijk}\\
\tag{M3}
Y_{ijk} &= a +a_j+b t_k+b_j t_k+c t_k^2+c_j t_k^2+\epsilon_{ijk}
\end{align}
Model (M0) is the simplest: the expected log ratio $Y_{ijk}$ is modeled by a
straight line, the slope $b$ of which does not depend on dilution. In model
(M1), the slopes $b+b_j$ may depend on dilution. In the actual fit,
the slopes are found to decrease as the initial proportion of RL increases. 
This is coherent with Figure \ref{fig1}. 
However, it hides the relevance of the quadratic models. 
Model (M3) is the complete log-quadratic model: both the slopes
$b+b_j$ and the curvatures $c+c_j$ may depend on dilution. Models
(M2b) and (M2c) are embedded into (M3): in model (M2b)
the slope does not depend on dilution and
in (M2c) the curvature does not depend on
dilution. 

For all four models, the linear fit was computed, then pairs of
embedded models were tested by the Fisher test of 
analysis of variance. The results are presented in Table
\ref{tests_models}: the degrees of freedom df, the Fisher test
statistic F, and the significance p-values are given. The conclusions
are the following. The first three comparisons are significant,
ie the bigger model is better than the embedded one. The
(M3) \emph{vs.} (M2c) comparison is not. The conclusion of the four comparisons
is that the best fitted model is (M2c).
This indicates that if curvatures are
included in the model, the slopes do not significantly 
depend on dilution. This is coherent with the theoretical derivation of
the log-quadratic model (\ref{Rlq}). In model (M2c) the estimated
slope is $\hat{b}=6.1\times 10^{-3}$ and 
$95\%$ confidence interval on $b$ 
is $[5.6\times 10^{-3};6.6\times 10^{-3}]$. Recall from (\ref{Rlq})
that the slope $b$ of (M2c) 
should be understood as the difference between the slopes of models 
(\ref{modelseparate}) for RL and THP-1. From the two previous
sections, the estimated difference is $1.01\times 10^{-2}$, which is
above the confidence interval on $b$ in (M2c).
\begin{table}
\begin{center}
\begin{tabular}{lccc}
\hline
Embedded models &df&F&p-value\\\hline
(M3) \emph{vs.} (M2b)&$(2,291)$&$24.67$&$1.3\times 10^{-10}$\\
(M2b) \emph{vs.} (M1)&$(1,293)$&$22.8$&$8\times 10^{-39}$\\
(M1) \emph{vs.} (M0)&$(2,294)$&$118$&$1.8\times 10^{-38}$\\
(M3) \emph{vs.} (M2c)&$(2,291)$&$2.68$&$0.0705$\\\hline
\end{tabular}
\caption{Tests of embedded models for the log ratio of RL vs THP-1
  cell counts.}
\label{tests_models}
\end{center}
\end{table}

Next, we tested the three pairwise differences of curvatures $c_j$ in
the accepted model (M2c). The results are presented in Table
\ref{tests_c}: the value of the
Student test statistic t, and the p-value are given (degrees of
freedom: 293).
All three
differences are significant. The estimated values of the curvatures $c+c_j$'s are
given in Table \ref{estimated_curvatures}. It turns out that
$c_1>c_2>c_3$. So the curvature $c+c_j$ decreases as the
initial proportion of RL increases. This is a similar phenomenon as
was observed on Figure (\ref{fig1}). Indeed, when curvatures are
neglected (model (M1)), the slopes were found to be decreasing
as the initial proportion of RL increases. The theoretical
explanation is given by the mathematical model. Model (M1) amounts to
keeping only the first term in the Taylor expansion (\ref{cgf}). Since
the next term is positive, the adjusted values of the slopes
increase with time. Model (M2c) considers the first two terms in
(\ref{cgf}), neglecting the third one. If that neglected term is
positive, then the same effect will occur: ajusted values of
curvatures increase with time. The third term is proportional to
the skewness of $B$. Observing decreasing curvatures as in Table
\ref{estimated_curvatures} is an indication that the skewness of $B$
may be positive, where $B$ is the (random) difference in growth rate
between RL and THP-1. Koutsoumanis and Lianou\cite{KoutsoumanisLianou13}
have proposed a logistic distribution as a model for growth rate
variability. That distribution has null skewness. We conjecture that
distributions with positive skewness provide better models for
variable growth rates.
\begin{table}
\begin{center}
\begin{tabular}{ccc}\hline
null hypothesis     & t&p-value \\\hline
$c_1=c_2$& $ 2.4$&$0.016$\\
$c_1=c_3$& $17.9$&$6.3\times 10^{-49}$\\
$c_2=c_3$& $16.1$& $4.5\times 10^{-42}$\\\hline
\end{tabular}
\caption{Pairwise tests for differences in curvatures $c_j$ 
in model (M2c).}
\label{tests_c}
\end{center}
\end{table}

\begin{table}
\begin{center}
\begin{tabular}{lc}
\hline
 Dilution $j$ & $c+c_j$\\\hline
$j=1$ ($0.5\%$) & $5.3\times 10^{-6}$\\
$j=2$ ($1\%$)& $4.96\times 10^{-6}$\\
$j=3$ ($5\%$)& $1.6\times 10^{-6}$\\\hline
\end{tabular}
\caption{Estimated curvatures $c+c_j$ in model (M2c).}
\label{estimated_curvatures}
\end{center}
\end{table}
\vskip 5mm\noindent
\section*{Conclusion}
That unchecked populations grow exponentially fast is a well known
fact, backed up by countless experiments, which validate the mathematical
theory of branching processes\cite{KimmelAxelrod02}. To any clone
stemming from a single cell can be associated 
an \emph{exponential growth rate}, also
called Malthusian parameter. It can be seen as the slope over a large
interval of time, of the line fitting logarithms of the number of
cells against time. What is questioned here, is the idea that
clones stemming from different cells in a given strain, should have
the same growth rate. Unlike Koutsoumanis and
Lianou\cite{KoutsoumanisLianou13} or Tomelleri et
al \cite{Tomellerietal08}, we do not provide direct evidence 
for the intrinsic variablility of growth rates, but instead an
indirect proof, coming from a simultaneous 
growth experiment\cite{Dykhuizen90}.

It consisted in growing in the same vessels two cancer cell lines,
RL\cite{Beckwithetal90} and 
THP-1 \cite{Tsuchiyaetal80}. If there existed a single growth rate for
all RL clones, and another for all THP-1 clones, then the ratio should
grow exponentially, the rate being the difference of
the two growth rates. In that case, the growth
rate of the ratio should not depend on the initial proportion of RL
vs. THP-1. Our
observations disproved this: the growth rate of the ratio was found to
increase as the initial proportion of RL decreased (Figure \ref{fig1}).
Assuming that growth rates may vary among clones provides
both intuitive, and theoretical explanations.

The intuitive explanation is the following. Consider a growth rate
as attached to each cell of a given clone. If clones may grow at
different rates, the proportion of cells in faster growing clones
will gradually increase. In other words, the distribution of growth
rates at increasing times will be shifted toward larger values. This
explains why, when the initial proportion of RL cells is $0.5\%$,
at the time it reaches $5\%$, the population of RL contains
more fast breeders than at time $0$. Therefore, the (apparent) growth
rate for an initial proportion of $0.5\%$ is larger than for an
initial proportion of $5\%$.

The theoretical explanation is the following. If growth rates of different
clones are considered as independent
random variables with a positive variance, then the model fitting the
logarithms of cell numbers against time must contain a quadratic
term, proportional to the variance of growth rates: 
variable growth rates imply that higher order terms
must be added to the
classical log-linear model. Now if a population
grows according to a log-quadratic model, and a log-linear model is
fitted instead, then the estimated slope over an interval of time
should increase as the interval moves to the right 
(see Figure \ref{fig2}). This will overestimate the mean growth
rate. 

To validate our theoretical explanation, we had to compare the fits of
the log-linear and log-quadratic 
models on our experimental data. On separate growth
data, the log-linear model was better than the log-quadratic model on
RL, the contrary was true for THP-1. On simultaneous growth data, the
log-quadratic clearly provided a better fit. This may seem
paradoxical: indeed, the same model should be adopted for separate and
simultaneous growths. The explanation of this apparent contradiction is
statistical. The estimated curvature terms are in all cases smaller
by several orders of magnitude than the estimated slopes. Therefore,
the log-linear and log-quadratic models can hardly be distinguished
when the cell counts range over several orders of magnitude, as in separate
growths. This cannot be the case on simultaneous growth data, where
the ratios range from a few percents to $100\%$. 
We believe that, on a separate growth
experiment, if more values were collected at the beginning, then the
log-quadratic model would  provide a better fit.
 
There remains the issue of a probabilistic model to be fit on variable
growth rates. Our derivation of the cumulant generating function shows
that classical models of positive random variables, such as
Gamma, Log-normal, or Logistic distributions\cite{KoutsoumanisLianou13} 
cannot be used here. Indeed their exponentially decaying tail implies
that the \emph{equivalent growth rate} would become infinite at finite
time, which is not realistic. So a truncated model would have to be
used instead. In any case, there would remain to adjust the chosen
distribution to actual data. Ideally, these data should be collected
from the observation of colonial growth of individual cells, such as
reported by Koutsoumanis and Lianou for 
\emph{Salmonella enterica}\cite{KoutsoumanisLianou13}, or Tomelleri et
al\cite{Tomellerietal08} on leukemia cells. This will be the
object of future work.
\vskip 5mm\noindent
\section*{Acknowledgements}
This work was supported by Laboratoire d'Excellence TOUCAN (Toulouse
cancer).
The authors are indebted to the anonymous referees for important
suggestions.
\section*{Disclosure}
The authors report no conflicts of interest in this work.
%
%\bibliographystyle{unsrt}
%\bibliography{/home/ycart/recherche/IS/IS.bib}






% \end{thebibliography}

\end{document}